\documentclass[onecolumn,,amsmath,amssymb]{revtex4}

\pdfoutput=1

\usepackage{graphicx}% Include figure files
\usepackage{dcolumn}% Align table columns on decimal point
\usepackage{bm}% bold math
\newcommand{\etal}{\emph{et al.}}
\newcommand{\be}{\begin{equation}}
\newcommand{\ee}{\end{equation}}
\newcommand{\bfig}{\begin{figure}}
\newcommand{\efig}{\end{figure}}

\begin{document}      
\title{Large discrete jumps observed in the transition between Chern states in a ferromagnetic Topological Insulator 
} 

\author{Minhao Liu$^{1,\dagger}$, Wudi Wang$^1$, Anthony R. Richardella$^2$, Abhinav Kandala$^2$, Jian Li$^1$, Ali Yazdani$^1$, Nitin Samarth$^2$, N. P. Ong$^{1,\dagger}$}
\affiliation{
$^1$Department of Physics, Princeton University, Princeton, NJ 08544\\
$^2$Department of Physics, the Pennsylvania State University, University Park PA 16802
}

\date{\today}      
\begin{abstract}
\end{abstract}
 
 % Produces the title
\maketitle      
{\bf Abstract\\
A striking prediction in topological insulators is the appearance of the quantized Hall resistance when the surface states are magnetized. The surface Dirac states become gapped everywhere on the surface, but chiral edge states remain on the edges. In an applied current, the edge states produce a quantized Hall resistance that equals the Chern number ${\cal C} = \pm 1$ (in natural units), even in zero magnetic field. This quantum anomalous Hall effect was observed by Chang~\etal. With reversal of the magnetic field, the system is trapped in a metastable state because of magnetic anisotropy. Here we investigate how the system escapes the metastable state at low temperatures (10--200 mK). When the dissipation (measured by the longitudinal resistance) is ultralow, we find that the system escapes by making a few, very rapid transitions, as detected by large jumps in the Hall and longitudinal resistances. Using the field at which the initial jump occurs to estimate the escape rate, we find that raising the temperature strongly suppresses the rate. From a detailed map of the resistance vs. gate voltage and temperature, we show that dissipation strongly affects the escape rate. We compare the observations with dissipative quantum tunneling predictions. In the ultralow dissipation regime, there exist two temperature scales $T_1\sim$ 70 mK and $T_2\sim$145 mK between which jumps can be observed. The jumps display a spatial correlation that extends over a large fraction of the sample. 
}

\section{INTRODUCTION}\label{sec:intro}

In the semiconductors known as topological insulators~\cite{FuKane,QiHughes,Kane,Qi} (exemplified by Bi$_2$Se$_3$, Bi$_2$Te$_3$ and Bi$_2$Te$_2$Se), strong spin-orbit interaction together with band inversion lead to electronic surface states which display a Dirac-like linear dispersion. On each surface, the Dirac states have only one spin degree of freedom, with the spin locked transverse to the momentum. The node of the Dirac cone is protected against gap formation if time-reversal symmetry (TRS) prevails (this constrains the nodes to be pinned at time-reversal invariant momenta). 

The unusual properties of the surface Dirac states have been intensively investigated by angle-resolved photoemission (ARPES), scanning tunneling microscopy (STM) and transport experiments. A natural question is what happens when TRS is broken by inducing a magnetization at the surface? Theory predicts that breaking of TRS opens a gap at the Dirac node throughout the surface ``bulk''~\cite{FuKane,QiHughes,Kane,Qi}. However, there remains a conducting edge state that runs around the perimeter of the sample. The edge state is chiral (it is either right or left-moving depending on the magnetization vector) and dissipationless (backscattering of the electrons is prohibited). As in the quantum Hall effect (QHE), the edge state displays a Hall resistance that is rigorously quantized ($R_{yx} = h/e^2$ where $h$ is Planck's constant and $e$ the electron charge) while the longitudinal resistance $R_{xx}$ vanishes~\cite{FuKane,QiHughes,Kane,Qi,SCZhang}. However, the quantization of $R_{yx}$ occurs even if the external magnetic field $\bf H$ is zero. This phenomenon is known as the quantum anomalous Hall (QAH) effect.

The quantization of $R_{yx}$ in zero field was first established experimentally by Chang \etal~ using ultrathin films of Cr-doped (Bi,Sb)$_2$Te$_3$~\cite{Xue}. Similar results in related TI films were subsequently reported by several groups~\cite{Checkelsky,KangWang,Moses,Goldhaber,Nitin,WeiLi,Yayu}. The topological nature of the Hall quantization produces a hysteretic loop in $R_{yx}$ that reflects the magnetic hysteresis when $\bf H$ (the applied magnetic field) is slowly cycled beyond the coercive field $H_c$. If the system is prepared with magnetization $\bf M\parallel H\parallel -{\bf \hat{z}}$, the chiral edge modes lead to $R_{yx} = -h/e^2$ = -25.812 k$\Omega$. This state is characterized by a Chern number ${\cal C} = -1$. As $H$ changes sign, the system becomes metastable; to exit the metastable state, it undergoes a very sharp transition to the state with ${\cal C} = 1$ and $R_{yx}=+h/e^2$. Here, we focus on the nature of this transition at millikelvin temperatures $T$. In conventional magnets, the hysteretic transition reflects the gradual diffusive motion of domain walls. Surprisingly, the transition actually proceeds by large, discrete jumps in $R_{yx}$. We find that slightly increasing $T$ suppresses the jump probability. A detailed investigation of how the jump probability varies with $H$ and $T$ suggests that the jumps reflect quantum tunneling events in the presence of dissipation.

\section{RESULTS}
\subsection{Characterization and benchmarks}\label{sec:bench}
\begin{figure*}[t]
\includegraphics[width=14cm]{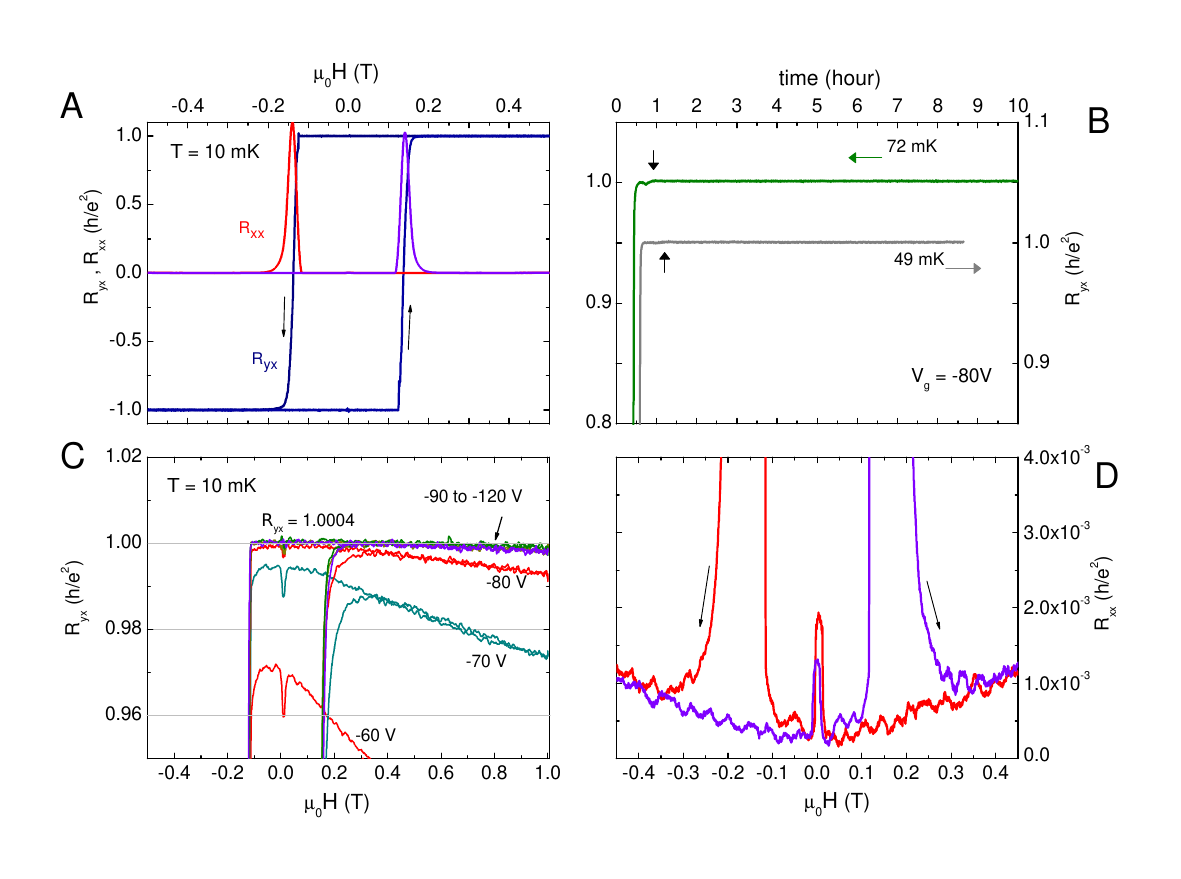}
\caption{\label{figQAHE}
Quantization of the Hall resistance in a 10 nm-thick film of Cr-doped (Bi,Sb)$_2$Te$_3$ at mK temperatures. Panel (A) shows the Hall plateaus in the Hall resistance $R_{yx}$ at 10 mK (blue curves). The transition between Chern states at the coercive field $H_c$ ($\sim$ 0.14 T) is nearly vertical. The longitudinal resistance $R_{xx}$ (red curves) displays ultralow dissipation, apart from the sharp peaks at $H_c$. Panel (B) tests the stability of the Chern state $R_{yx} = h/e^2$ over an extended period (8-9 hrs), with $H$ fixed at zero starting at time indicated by vertical arrows. The two traces at 49 mK (grey trace) and 72 mK (green) are shown displaced for clarity. In Panel (C), the expanded scale shows that $R_{yx}$ deviates from $h/e^2$ by a few parts in $10^4$ in the optimal gating window $-110 <V_g < -90$. In Panel (D), the expanded scale shows that $R_{xx}$ drops to $<3\times 10^{-4} h/e^2$ near $H=0$. The sharp spike at $H=0$ reflects quenching of superconductivity in the In solder used in the contacts. $R_{xx}$ is thermally activated with a gap $\Delta_R\sim$ 190 mK.
}
\end{figure*}

In the experiment, we used samples of Cr-doped (Bi,Sb)$_2$Te$_3$ grown on a SrTiO$_3$ substrate by molecular beam epitaxy to a thickness of 10 nm and cut into Hall bars (0.5 mm $\times$ 1 mm) (see Methods and Ref.~\cite{Nitin}). As grown, the surface states are $n$-doped with the chemical potential lying high above the surface Dirac node. A large, negative back-gate voltage $V_g$ = -80 V is required to lower $\mu$ to the surface Dirac node. Within the optimal gate window -120 $< V_g <$-80 V (the limits are nominal; they change by $\pm 10\%$ in successive experiments), the Hall resistance $R_{yx}$ attains near-ideal quantization at temperature $T$ = 10 mK. Figure \ref{figQAHE}A shows that, at $H_c \sim 0.14$ T, $R_{yx}$ undergoes narrow transitions between -1 and +1, with narrow peaks in the resistance $R_{xx}$ of width $\Delta H\sim$ 20 mT (hereafter, we quote resistances in units of $h/e^2$).

An important question is the stability of the quantized values $R_{yx}=\pm 1$. If the magnetization is actually not spontaneous, $R_{yx}$ should relax away from the quantized values when $H$ is returned to 0, whereas a truly magnetized ground state with ${\cal C} = \pm 1$ should display no sign of relaxation. To test this, we prepared the state in positive field $\bf H\parallel \hat{z}$. We then returned $H\to 0$ and monitored $R_{yx}$ for a period 8-9 hrs. As shown in Fig. \ref{figQAHE}B, we do not detect any sign of relaxation at 49 mK and 72 mK for $V_g$ = -80 V. To us this shows that, at the Hall plateaus, the system is in the true ground state with ${\cal C} = \pm 1$. Moreover, it remains stable in the limit $H\to 0$ at least to 8-9 hrs. 

The expanded scale in Fig. \ref{figQAHE}C shows that $R_{yx}$ deviates from 1.0000 by $\sim$4 parts in 10$^4$ within the optimal gating window. Figure \ref{figQAHE}D shows in expanded scale the behavior of $R_{xx}$ away from the peaks. Significantly, $R_{xx}$ at $H$=0 falls to values $<3\times 10^{-4}$ at 10 mK consistent with ultralow dissipation. These benchmarks are comparable to the best achieved to date~\cite{Moses,Goldhaber}. 

\begin{figure}
\includegraphics[width=8cm]{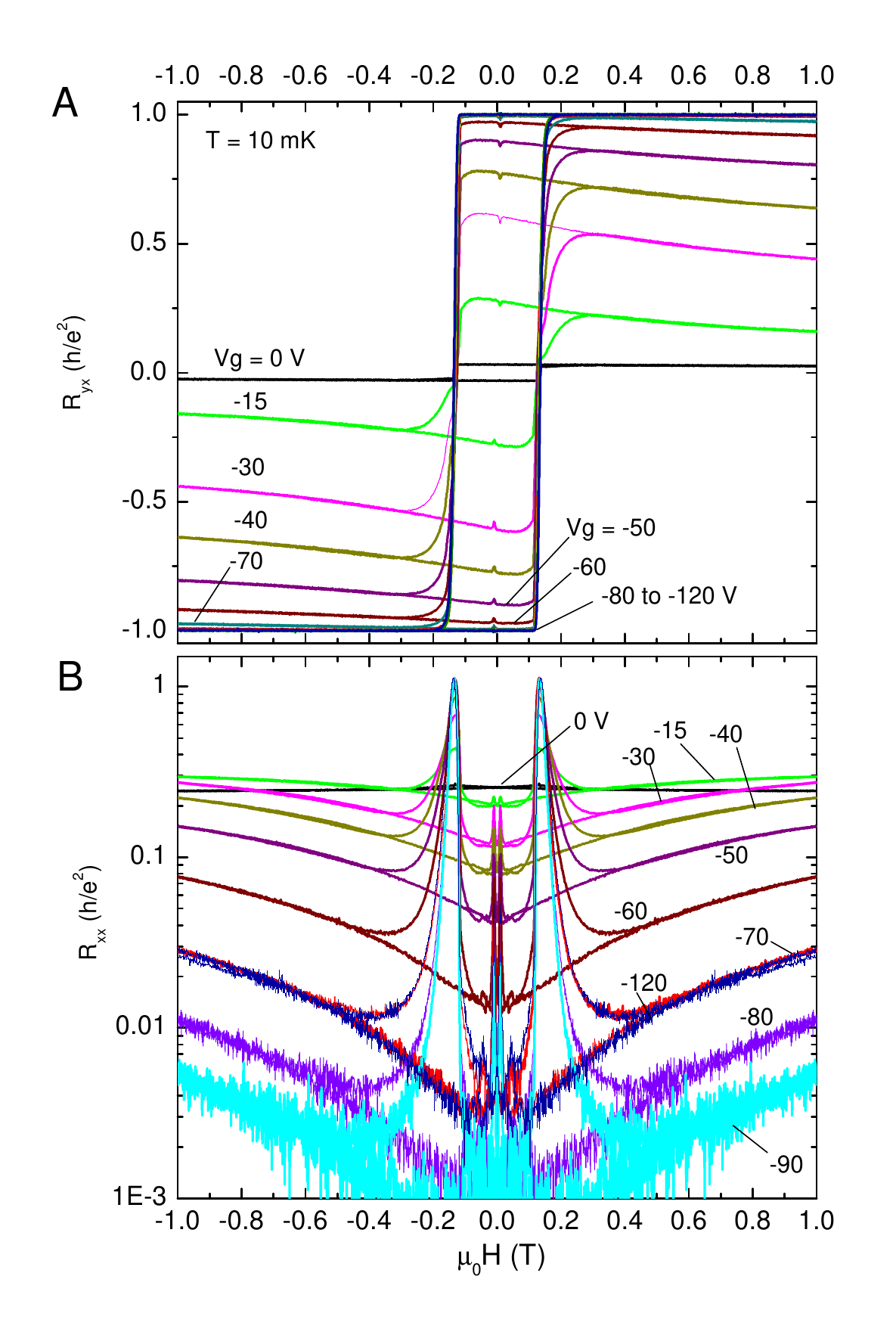}
\caption{\label{figGate}
The dependences of $R_{yx}$ and $R_{xx}$ on gate voltage $V_g$ at $T$ = 10 mK. Panel (A) shows traces of $R_{yx}$ vs. $H$ at selected $V_g$ from 0 to -120 V. The negative slopes at large $H$ show that the bulk carriers which appear when $|V_g|<$ 40 V are $n$-type. Panel (B) plots $R_{xx}$ for the same values of $V_g$ in semilog scale. The coercive field $H_c$ determined by the sharp peaks in $R_{xx}$ are nearly independent of $V_g$. 
}
\end{figure}

To provide a broader view of the transport behavior at 10 mK, we show in Fig. \ref{figGate} the curves of $R_{yx}$ and $R_{xx}$ vs. $H$, with $V_g$ set at values from 0 to -120 V. The Hall curves $R_{yx}$ (Fig. \ref{figGate}A) remain very close to 1.0000 in the optimal gate window, but start to fall monotonically as $|V_G|$ decreases, reaching 0.03 when $V_g$ = 0 (surface states strongly $n$-doped). Throughout the gate interval, -70 to 0 V, the sign of the $n$-type carriers imparts a gentle slope to $R_{yx}$ for fields $|H|>H_c$. Turning to the longitudinal resistance, we show in semilog scale the curves of $R_{xx}$ vs. $H$ for the same gate voltage values (Fig. \ref{figGate}B). As noted above, $R_{xx}$ falls to values as low as $3\times 10^{-4}$ (in terms of $h/e^2$) within the field window $|H|<H_c$ (see curve at $V_g$ = -90 V). When $|H|$ is increased beyond $H_c$, the dissipation increases exponentially. As we tune $V_g$ outside the optimal gate window, the curves of $R_{xx}$ also increase exponentially, eventually reaching $\sim$0.2 when $V_g$ approaches 0. The pattern of an exponential increase in dissipation when either $V_g$ is tuned outside its optimal window, or $|H|$ is increased above the coercive field, implies the existence of a small energy gap that ``protects'' the dissipationless state with accurately quantized $R_{yx}$. We estimate the activation gap below (see Sec. \ref{sec:gap}).

%%%%%%%%%%%%%%%%%%%%%%%%%%
%%%%%%%%%%%%%%%%%%%%%%%%%%
%%%%%%%%%%%%%%%%%%%%%%%%%%

\subsection{Jumps in $R_{xx}$ and $R_{yx}$}\label{sec:jumps}

\begin{figure*}[t]
\includegraphics[width=16cm]{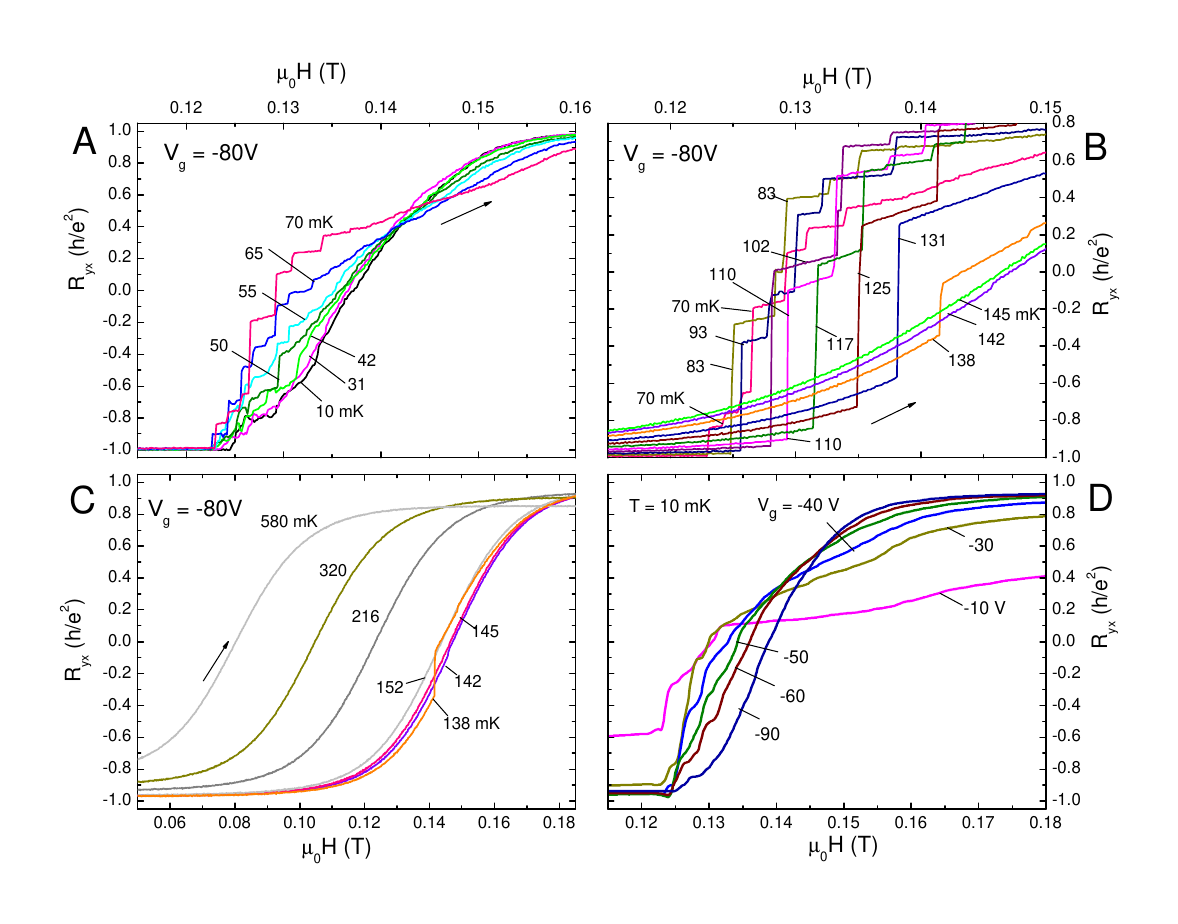}
\caption{\label{figRyxvsT}
Jumps in $R_{yx}$ at the transition between Chern states. Panel (A) shows traces of $R_{yx}$ at selected $T$ (10$\to$70 mK) as $H$ is slowly swept (10 mT/min) past $H_c$ as indicated by the arrows. The transition between Chern states $C\, =\, \pm1$ is nominally smooth below 40 mK. Starting near $T_1\sim$ 65 mK, small vertical jumps appear near the start of the transition. Panel (B) shows $R_{yx}$ vs. $H$ in the important interval 70 mK $\le T\le$ 145 mK in which large jumps are clustered. Starting at 83 mK, the system ``escapes'' the $C=-1$ state by a very large initial jump, followed by a cascade of smaller ones. Focussing on the initial jump, we see that field $H_J(T)$ that triggers the jump (kink feature) increases steadily as $T$ rises to 142 mK. Above 131 mK, the jump magnitude $\Delta R_{yx}$ sharply decreases becoming unresolved above $T_2\sim$ 145 mK. Panel (C) shows that, from $T_2$ to 580 mK, $R_{yx}$ changes smoothly over the (now broadened) transition. Panel (D) shows how the $R_{yx}$ curves change with $V_g$ at 10 mK. At optimal gating ($V_g$ = -90 V), the transition is smooth, but as $V_g$ is increased to -10 V, the jumps re-appear, implying that a small amount of dissipation is necessary to seed the jump at the lowest $T$. The experimental time constants are 1s in all panels except D where it is 5 s (see Supplement). 
}
\end{figure*}

The main results -- observation of large jumps in $R_{xx}$ and $R_{yx}$ -- are described in this subsection.
At each $T$, we prepare the system with Chern number ${\cal C} =-1$ ($\bf H\parallel -\hat{z}$). To investigate the transition ${\cal C}:-1 \to 1$, we scan $H$ very slowly (1-10 mT/min) across $H_c$ (left to right in all figures). Starting at our lowest $T$ (10--50 mK), $R_{yx}$ displays a nominally smooth profile as it changes from -1 to 1 (Fig. \ref{figRyxvsT}A). The smooth variation, which has a profile that is $T$ independent in width up to 580 mK, is likely associated with conventional domain-wall motion.

In the interval 50-65 mK, small precursor steps become discernible above an onset field $H_s$. At the characteristic temperature $T_1\sim$ 65 mK, we observe the appearance of remarkably large jumps with magnitudes $\Delta R_{yx}\sim 1$. Within the interval $T_1\to T_2$ = 145 mK, the jumps appear in a stochastic, yet systematic pattern as shown in Fig. \ref{figRyxvsT}B. Their occurrence is independent of the adopted sweep rate (1-10 mT/min). We focus on the first jump, which is always the largest (in over 50 scans recorded). As $T$ rises above $T_1$, its magnitude $\Delta R_{yx}$ initially increases (curves at 83$\to$ 131 mK). Above 138 mK, $\Delta R_{yx}$ rapidly diminishes to become unresolved. As shown in Fig. \ref{figRyxvsT}C, all curves above $T_2$ are smooth and devoid of jumps. Interestingly, even though jumps are not observed at 10 mK in the ultra-low dissipation state, they re-appear when we tune $V_g$ to exit the optimal window. In Fig. \ref{figRyxvsT}D, $R_{yx}$ at -90 V is smooth, but jumps appear when $V_g$ is raised above -40 V. The jumps are reproducibly observed even after repeatedly warming to 300 K. They are distinct from magnetization steps observed in manganites at $T$ as high as 4 K~\cite{Schiffer}.

\begin{figure*}[t]
\includegraphics[width=16cm]{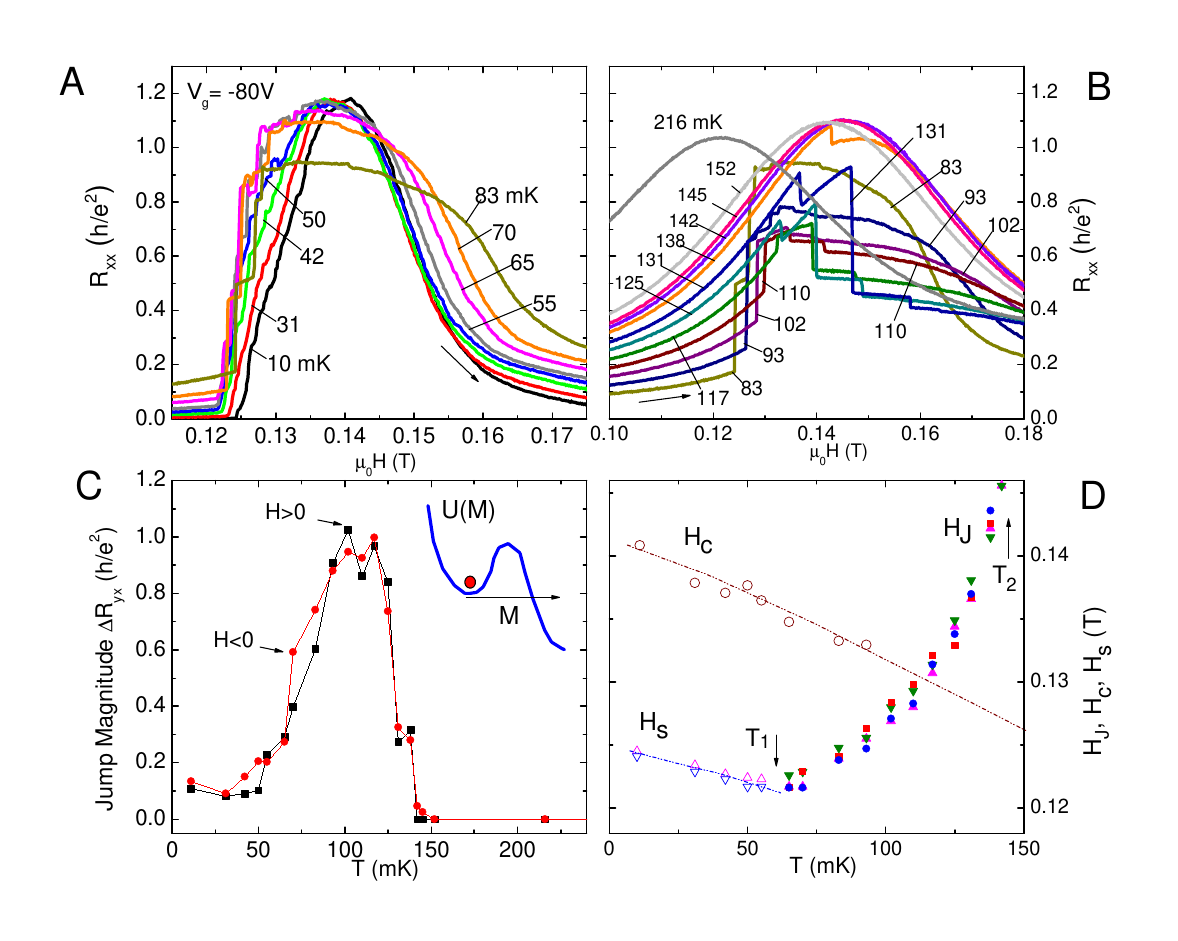}
\caption{\label{figHJ}
Profiles of resistance $R_{xx}$ vs. $H$ at selected $T$ as $H$ is slowly swept at 1 mT/min (arrows). Below 50 mK (Panel (A)), the profile is nominally smooth and asymmetric, whereas small vertical jumps appear near the leading edge as $T$ rises above $T_1$. In the interval 83$\to$ 216 mK (Panel (B)), the smooth profile is dissected by large jumps $\Delta R_{xx}$ that are positive (upwards) near the leading edge and negative near the trailing edge. Above $T_2$, the profile is again smooth but twice as broad as at 10 mK. In Panel (C), we plot the distribution of the initial jump magnitude $\Delta R_{yx}$ vs $T$ within the optimal gating window. Above $T_1$, $\Delta R_{yx}$ initially rises to a value $\sim h/e^2$, then falls steeply to zero at $T_2$. Jumps are not resolved above $T_2$. The inset is a sketch of the system (red circle) trapped in a potential $U(M)$. Panel (D) plots the escape field $H_J(T)$ of the initial jump vs. $T$ (solid symbols) inferred from $R_{yx}$ and $R_{xx}$ from both sweep-up and -down runs. Below $T_1$, $H_J(T)$ terminates at the curve of $H_s(T)$ (the onset field for small, precursor steps plotted as open triangles), which shares the same slope as the coercive field $H_c(T)$ (open circles). $H_c$ is defined by the maximum in $R_{xx}$ (when the curve is smooth). $H_c$ is undefined between 100 mK and $T_2$, but it smoothly extends to measurements above $T_2$. 
}
\end{figure*}

The jumps are also prominent in the dissipative channel (curves of $R_{xx}$). At low $T$ (10-50 mK), $R_{xx}$ exhibits a nominally smooth, asymmetric profile (Fig. \ref{figHJ}A). Above $T_1$, large jumps appear at the leading (left) edge while the trailing edge remains smooth, as the width increases to 38 mT. Figure \ref{figHJ}B follows these curves above $T_2$ to 216 mK. As noted for $R_{yx}$, the escape field $H_J(T)$ of the initial jump increases with $T$. However, in the window 125-138 mK, the jumps now have opposite signs (positive at the leading edge). Jumps observed in $R_{yx}$ and $R_{xx}$ when $H$ is swept in the opposite direction are closely similar (Supplementary Figs. S2 and S3).

Figures \ref{figHJ}C and \ref{figHJ}D show how the two important quantities (the jump magnitude and the initial field) vary with $T$. As noted, the jump magnitude $\Delta R_{yx}$ is initially small within the precursor interval, but increases rapidly at $T_1$ to saturate at a plateau value $\sim$1 (Fig. \ref{figHJ}C). As $T$ is further raised above $T_2$, $\Delta R_{yx}$ collapses to a value below our resolution. Above $T_2$ (up to 500 mK), the Hall curves vary smoothly throughout the hysteresis loop. In contrast to the non-monotonicity of $\Delta R_{yx}(T)$, the escape field $H_J(T)$ increases steeply from $T_1$ to $T_2$ (solid symbols in Fig. \ref{figHJ}D). It is interesting that $H_J$ is still rising with steepening slope when the jump magnitude falls below resolution above $T_2$. For comparison, we also plot the gentle decrease of $H_c$ and $H_s$ (open symbols) over the same interval of $T$ (obtained from curves in Supplementary Fig. S4).

\subsection{Spatial Correlations}
\begin{figure*}[t]
\includegraphics[width=16cm]{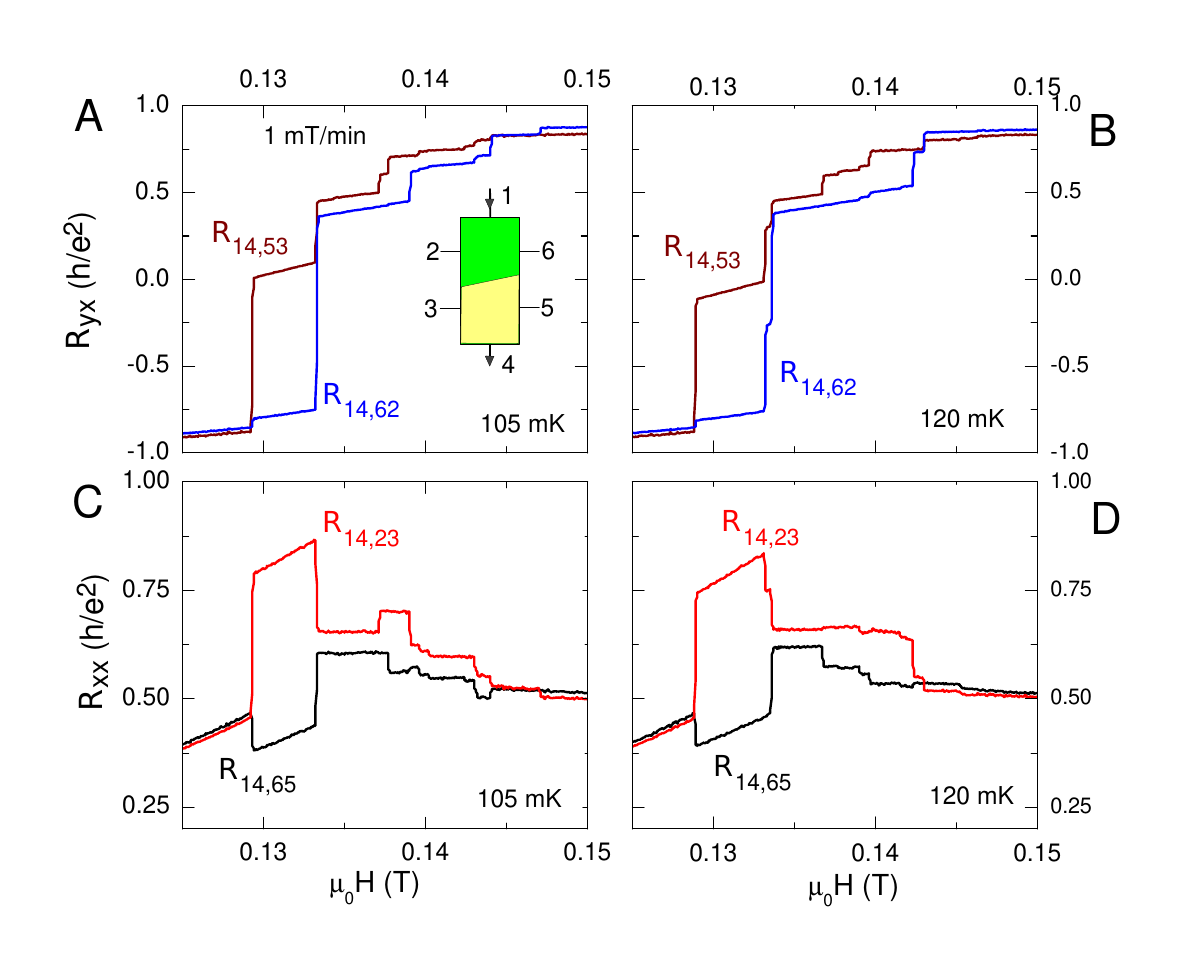}
\caption{\label{figCorr}
The jumps in $R_{yx}$ (top panels) and $R_{xx}$ (bottom) detected by simultaneous measurements of $R_{14,mn}$, as labelled [$I$ applied to contacts $(1,4)$ and voltage measured across $(m, n)$]. The field sweep rate is 1 T/min. At both 105 and 120 mK (Panels (A) and (B)), the initial jump strongly affects the downstream Hall signal ($R_{14,53}$), but barely changes the upstream pair ($R_{14,62}$). At the second jump, however, the upstream pair is strongly perturbed. The inset in Panel (A) shows the contacts, and the wall separating Chern state ${\cal C} = -1$ (green) from +1 (yellow) after the first jump occurs. In the resistance (bottom row), the first jump increases $R_{14,23}$ but decreases $R_{14,65}$ at both 105 and 120 mK (Panels (C) and (D)). The signs in both pairs become reversed at the second jump, also consistent with the upstream motion of the wall. 
}
\end{figure*}

An interesting way to study the jumps is provided by simultaneous measurements of $R_{ij}$ across well-separated pairs of voltage contacts. The resistance $R_{ij,mn}$ is defined as $V_{mn}/I_{ij}$ where $V_{mn}$ is the voltage measured across contacts $(m,n)$ with the current $I_{ij}$ injected at the source contact $i$ and drained at $j$ (see inset, Fig. \ref{figCorr}A). A typical pattern is that a jump in $R_{yx}$ measured across one pair of contacts barely affects the signal across a different pair, but at a slightly larger $H$ the reverse holds. 

Figures \ref{figCorr}A and \ref{figCorr}B show traces of $R_{14,53}$ and $R_{14,62}$ vs. $H$ at $T$= 105 mK and 120 mK, respectively. In both panels, the initial jump (at $H$ = 0.129 T) is large across the downstream Hall pair (5,3) while the upstream Hall pair (6,2) is barely affected. At a larger $H$ (0.133 T), the second jump is prominent in the upstream pair but much smaller in the downstream pair. Simultaneously, the resistance $R_{14,23}$ in the right pair (2,3) first increases and then decreases whereas in the left-pair (6,5), $R_{14,65}$ shows the opposite sequence (Figs. \ref{figCorr}C and \ref{figCorr}D).

\begin{figure}
\includegraphics[width=8cm]{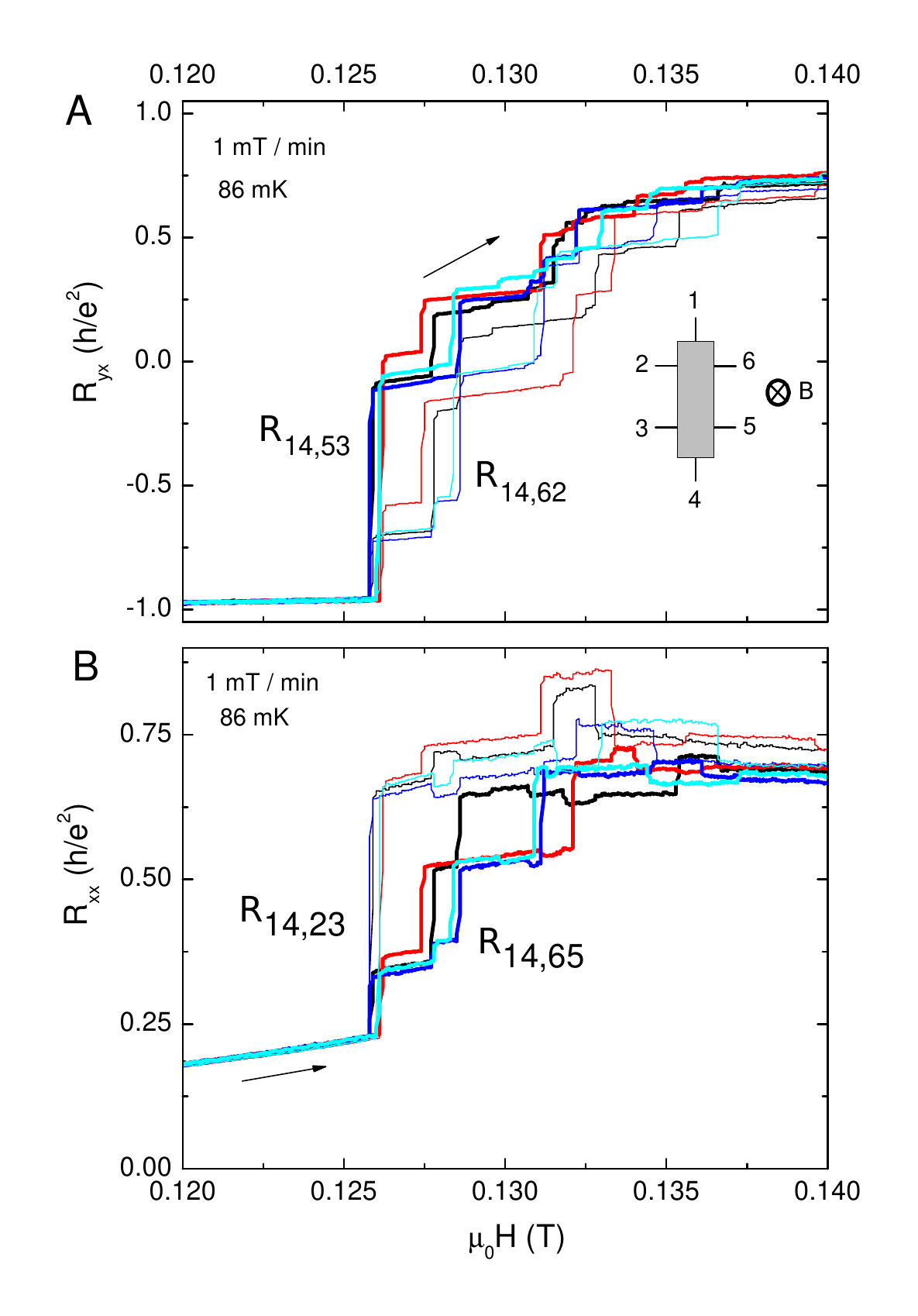}
\caption{\label{figMultiple}
Four traces of the resistances $R_{14,mn}$ measured in succession at 86 mK with $V_g$ = -80 V. For each trace, the sample is prepared in the ${\cal C}$ = -1 state and the transition is induced by sweeping $H$ slowly through $H_c$ at the rate 1 mT/min while monitoring the resistances $R_{14,mn}$. Panel (A) shows the upstream Hall resistance $R_{14,26}$ (light curves) and downstream Hall $R_{14,53}$ (bold). Panel (B) plots the corresponding longitudinal left-contact resistance $R_{14,65}$ (bold curves) and right-contact resistance $R_{14,23}$ (light). Although the jumps are stochastic, the overall pattern for the first two jumps is fairly reproducible from one run to the next.
}
\end{figure}

Despite the stochastic nature of the jumps, the overall pattern is systematic and reproducible, especially for the first two jumps in a trace. In Fig. \ref{figMultiple} we plot four successive scans of the traces taken successively with $T$ fixed at 86 mK and $V_g = -80 V$. The sample is prepared in the same way in each scan with the sweep rate fixed at 1 mT/min. Panel A reports the Hall resistances (light curves for the upstream pair, bold curves for downstream). In panel B, we show the corresponding longitudinal resistance curves (bold curves for the left pair, light curves for the right). The first two jumps are reproducible although the jump magnitudes and $H_J$ can vary by 10$\%$. The variations from one jump to the next shown here are quite typical of the reproducibility across the whole temperature range investigated (10 mK to 580 mK).

\begin{figure}
\includegraphics[width=8cm]{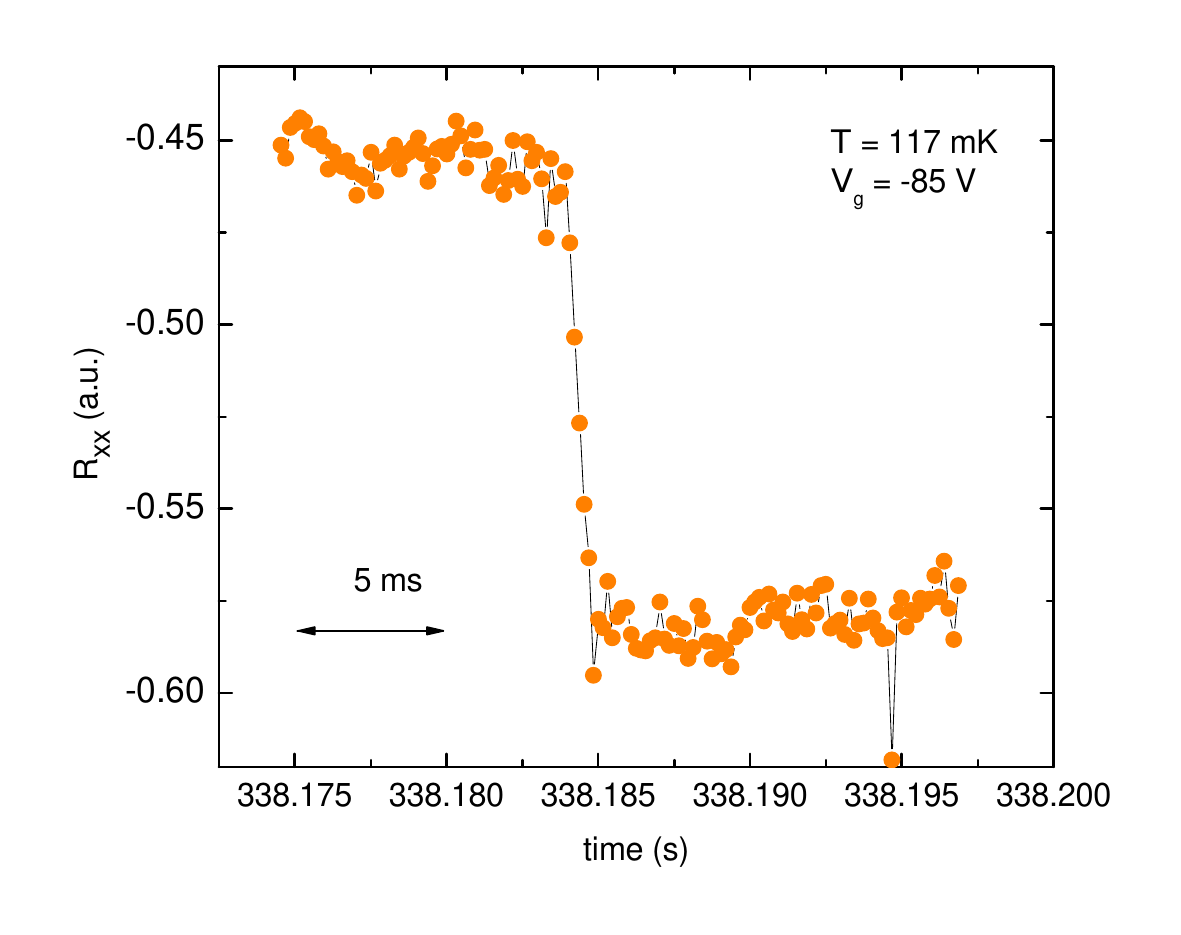}
\caption{\label{figTime}
Time trace of a jump event observed in $R_{xx}$ measured with the room-temperature filters in the circuit removed (but not the low-temperature filters) at $T$ = 117 mK and $V_g$ = -85 V. The jump occurs on a time scale $\Delta t$ that is shorter than 1 ms, the time constant of the remaining filter. The field sweep rate was 0.1 mT/min.}
\end{figure}

How fast are the jumps? To minimize extraneous perturbations, all the traces shown above were taken at very slow field-sweep rates (typically 1 mT/min). We have attempted to estimate the duration of the first jump $\Delta t$ by minimizing the capacitive loading in the measurement circuitry in the dilution refrigerator. To this end, we sequentially removed several stages of the filter elements in the circuit (which increased the noise in the traces), while slowing the field scan to 0.1 mT/min. The final trace in this process is measured with only the low-temperature filter present. The noisy trace (Fig. \ref{figTime}) shows that $\Delta t$ is shorter than 1 ms, the time constant of the remaining filter.

\subsection{Activation across a small gap }\label{sec:gap}

Extending the measurements over a broad range of $V_g$ and $T$, we find that the ultralow dissipation state is protected by a minigap $\Delta_R\sim$ 190 mK. Setting $V_g$ within the optimal gating window (-110 to -75 V) places $\mu$ inside $\Delta_R$ (these numbers can shift by $10\%$ between runs because of charge trapping at the interface). Once $\mu$ exits the minigap, dissipation rises exponentially. 

\begin{figure}
\includegraphics[width=8cm]{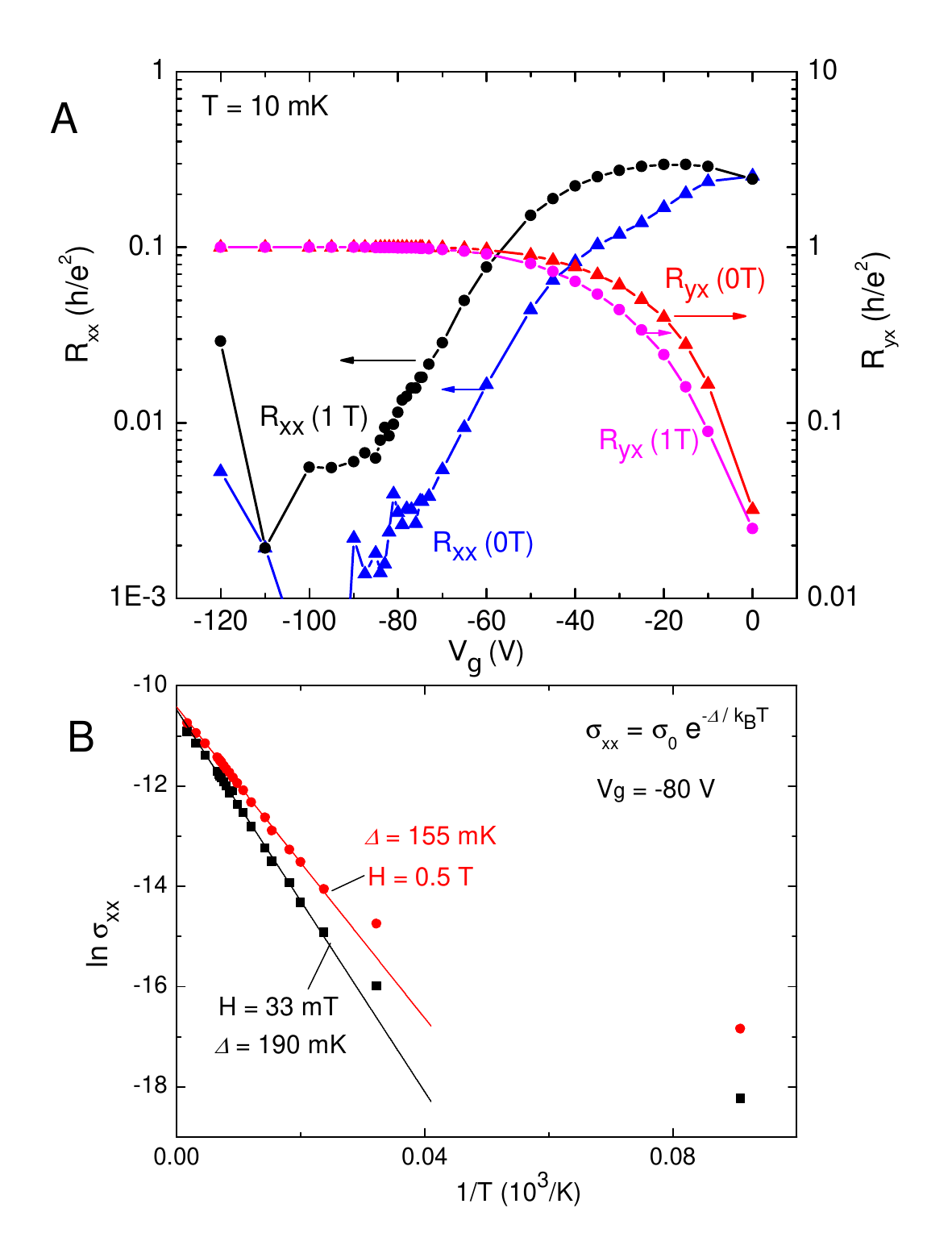}
\caption{\label{figR}
Semilog plot of $R_{yx}$ and $R_{xx}$ as a function of $V_g$ (Panel (A)), with $T$ fixed at 10 mK and $H$ fixed at either 0 (solid triangles) or at 1 T (solid circles). As $|V_g|$ decreases, the exponential increase in $R_{xx}$ (blue and black symbols) implies that $\mu$ is being raised above the mini-gap $\Delta_R$. Comparison between $R_{xx}$ at $H$ = 0 (blue) and 1 T (black) shows that a finite $H$ decreases $\Delta_R$. Similarly, the curves for $R_{yx}$ (red and magenta) show an exponential deviation from the quantum value $h/e^2$ as $|V_g|$ is decreased. The Hall curves persist longer at the quantized value than the resistance curves.
The semilog plot of $\sigma_{xx}$ vs. $1/T$ with $V_g$ = -80 V (Panel (B)) reveals an activated conductivity at both $H =$ 0.5 T and 33 mT. The straight-line fits yield values of $\Delta_R\sim$ 190 mK (at $H\simeq$ 0) and 155 mK (0.5 T). 
}
\end{figure}

Figure \ref{figR}A shows the changes to $R_{yx}$ and $R_{xx}$ as $V_g$ is varied from -120 V to 0, with $T$ fixed at 10 mK and $H$ fixed at $\sim$0 T (solid triangles) or 1 T (solid circles). The Hall curves (red and magenta symbols) show that, as $|V_g|$ is decreased, $R_{yx}$ stays close to the quantum value $h/e^2$ (the optimal gating window) until $|V_g|$ decreases below 70 V in zero $H$. Thereafter, $R_{yx}$ decreases exponentially to 0.03 as $V_g \to 0$. (The lower limit of the optimal window (-70 V in panel a) varies from run to run because of hystereses caused by charge trapping in the SrTiO$_3$ substrate.) The resistance $R_{xx}$ at $H$ = 0 (blue triangles) increases exponentially with a lower threshold than $R_{yx}$. A finite field (1 T, black circles) lowers the threshold significantly. We identify the exponential changes in both $R_{yx}$ and $R_{xx}$ with raising of $\mu$ above the upper limit of the gap $\Delta_R$ by gating.

The magnitude of the minigap $\Delta_R$ is determined by detailed measurement of the conductance $\sigma_{xx}$ vs. $T$. From the semilog plot of $\sigma_{xx}$ vs. $1/T$ in Fig. \ref{figR}B, we infer $\Delta_R\sim$ 190 mK for $H=$ 33 mT (black squares; $H$ is set slightly away from 0 to avoid the sharp feature arising from quenching of superconductivity in the In contacts). At 0.5 T, the equivalent plot gives a slightly smaller $\Delta_R$ (155 mK), consistent with the steeper increase in the curve with black symbols in Fig. \ref{figR}A. Both plots imply that the minigap decreases noticeably with increasing $H$. This is also consistent with the gradual increase of the background dissipation discussed in conjunction with Fig. \ref{figGate}B.

The changes to the transport curves induced by varying either $V_g$ or $T$ provide firm evidence that the ultralow dissipation state in the QAH state is protected by a minigap $\Delta_R\sim$ 190 mK. An exponential increase in the dissipation is observed when $T$ is increased or when we move $\mu$ out of the optimal window by gating. The size of the gap is also decreased by increasing $H$ beyond 0.5 T. The gradual increase of the background with $H$, together with the negative slope of $R_{yx}$ outside the coercive-field window, suggests that the bulk carrier population ($n$-type) is gradually increased by $H$ when $\mu$ falls outside the optimal window. To summarize the results, we plot schematically the behavior of the curves $\Delta_R$, $T_1$ and $T_2$ in the $V_g$-$T$ plane (Fig. \ref{figphase}).

\section{DISCUSSION}
\subsection{Metastability and the effect of $T$}\label{metastable}
The reversal of $\bf H$ from the direction in which the system is prepared ($\bf H\parallel -\hat{z}$) traps the system in a metastable state (spin $\downarrow$) until escape occurs near the coercive field. We have investigated the nature of the transition in the magnetic topological insulator at millikelvin temperatures. In a conventional ferromagnet escape from the metastable state corresponds to gradual motion of domain walls separating up and down domains, which allows the spin-$\uparrow$ domain to expand. This leads to the familiar smooth hysteretic $M$-$H$ loop. Here, however, the transition which involves both a change of the Chern number (${\cal C}=-1 \to 1$) and the average $\bf M$ displays a rich array of additional features, most notably the large jumps in $R_{yx}$ and $R_{xx}$ in the interval $T_1<T<T_2$ (Figs. \ref{figRyxvsT} and \ref{figHJ}). The jumps occur on time scales shorter than 1 ms. The first jump involves reversal of $\bf M$ over roughly half the sample. (As mentioned, we identify the smooth $T$-independent variation of $R_{yx}$ above $T_2$ with conventional domain-wall motion. Below $T_2$, the smooth variation persists as a background channel between the jumps.)

[In two recent investigations~\cite{WeiLi,Yayu} of the transition ${\cal C} = -1\to 1$, the variation of $R_{yx}$ is much broader than the ones observed here. Instead of jumps in $R_{yx}$, a shoulder at ${\cal C} =0$ (zero-Hall conductivity feature) is observed. Compared to the present sample (thickness 10 nm), the samples used are in the ultrathin limit (5 nm in Ref. \cite{WeiLi} and 6 nm in Ref. \cite{Yayu}) where hybridization between the states on the two surfaces is expected. Both the presence of zero-Hall shoulder and the absence of jumps suggest that these ultrathin samples are in a distinctly different regime of QAH behavior from the present sample. The degree of disorder appears to be higher in the ultrathin regime. For e.g., in the zero-Hall sample~\cite{Yayu}, the dissipation remains high at 50 mK ($R_{xx}\sim 4$ at $H$ = 0) while accurate Hall quantization is not attained ($R_{yx}\sim$ 0.9).]

We start by applying the standard Cahn-Hilliard nucleation theory to the expansion of a bubble of the true ground state ($\uparrow$) within a sample in the metastable state ($\downarrow$). The free energy of a 2D bubble of radius $R$ is
\be
F_b(R) = 2\pi R\sigma - \pi R^2\Delta F \quad (\Delta F = F_\downarrow - F_\uparrow),
\label{eq:Fb}
\ee
where $\Delta F>0$ is the free energy gain in the bulk of the bubble. The surface tension $\sigma$ which expresses the cost of the wall is proportional to the exchange energy. When $R$ exceeds the critical radius $R_c = \sigma/\Delta F$, the bubble expands without limit (if dissipation is absent), happily paying the cost of the wall to gain bulk energy. Conversely, if $R<R_c$, the bubble shrinks to zero. At the critical radius, $F_b$ has the critical value ${F_b}^c = \pi\sigma^2/\Delta F$. For a purely classical  (thermally activated) escape process, the lifetime of the metastable state is 
\be
\tau = \tau_0 \exp\left(\frac{{F_b}^c}{k_BT}\right),
\ee
where $1/\tau_0$ is an attempt frequency. Clearly, if $\sigma$ and $\Delta F$ are both $T$ independent (which we argue is the case here at all $T<$ 0.5 K), the liftetime of the metastable state decreases exponentially with increasing $T$.

Trapping of the system in a metastable state can be represented as a particle located in a potential well $U(\bf M)$ where $\bf M$ is a collective coordinate representing the local magnetization $\bf M(r)$ (see sketch in Fig. \ref{figHJ}C, inset). After the system is prepared with $\bf H\parallel-\hat{z}$ at temperature $T\sim$ 100 mK, it remains trapped for extremely long times if $H$ is tuned to a positive, but small, value $0<H<H_s(T)$ ($\simeq$0.122 T at 70 mK). As shown in Fig. \ref{figQAHE}B, the stability for $H$ =0 exceeds 8 hr at 10 mK. Incrementing $H$ above $H_s$ leads to a lowering of the barrier and a decrease in $t_E$. When $H$ equals $H_J(T)$, we infer that $t_E(T)$ falls inside the observation time window (1-3 min), so the first jump is observed. By measuring how $H_J(T)$ varies with $T$, we can deduce the effect of $T$ on the escape rate. 

The curve of $H_J(T)$ (Fig. \ref{figHJ}D) shows that $H_J$ increases steeply with $T$. This trend is counter-intuitive if we view the escape as a classical process. In such ``over the barrier'' processes, raising $T$ invariably speeds up the escape event. Quite generally, in the Langer formulation~\cite{Langer}, an escape occurs when a particular spin configuration creates a saddle point in the potential landscape $U({\bf M(r)})$ that breaches the barrier. As the probability for this to occur rises exponentially with $T$, raising $T$ should lead to a sharp reduction of $t_E$ (and hence a smaller $H_J$). This is opposite to the observed trend of $H_J(T)$. Contrary to the general expectation that ``escape'' should become easier, as more spin configurations become accessible at higher $T$, we find that escape is in fact more difficult. 

Next, we consider a hybrid classical-quantum process in which escape results from thermally-assisted tunneling. Absorption of bosons (phonons or spin waves) from the bath brings the system's energy closer to the top of the barrier thereby enhancing the tunneling rate. Clearly, raising $T$ increases the phonon or spin-wave population leading to an enhanced escape rate, again in conflict with $H_J(T)$.

\subsection{Dissipative quantum tunneling}\label{dissipative}
One way to obtain an escape rate that increases with decreasing $T$ is dissipative quantum tunneling. In this discussion, dissipation, as monitored by $R_{xx}$, plays an important role.

The quantum tunneling of a magnetization configuration out of a metastable state is closely related to the decay of the false vacuum investigated by Coleman and Callan~\cite{Coleman,Callan}. Via an instanton process (see Supplementary Sec. S3), the system tunnels ``under the barrier'' to form a small bubble of the true ground state. Thereafter, the bubble expands rapidly without limit (if dissipation is absent) in the classical Cahn-Hilliard process.

To incorporate dissipation, Caldeira and Leggett~\cite{Caldeira,Caldeira83} consider a single particle of mass $m$ in the tilted double-well potential $U(q)$, and described by the classical equation of motion $m\ddot{q}+\eta(\omega)\dot{q} + \partial U/\partial q=0$ ($q$ is the collective coordinate). Dissipation -- introduced by a linear coupling to many classical oscillators -- appears as the damping coefficient $\eta(\omega)$ which may depend on frequency $\omega$. They find that increasing the dissipation suppresses the tunneling probability $\Gamma$. In effect, each sampling (measurement) of the particle's position by an oscillator collapses its wave function; this resets the tunneling process. Whereas $\eta$ by itself decreases $\Gamma$, raising $T$ has the opposite effect~\cite{Weiss}. The opposing effects of dissipation and $T$ have been investigated experimentally~\cite{Washburn}. 

Within the optimal gating window in our experiment, $R_{xx}$ is thermally activated with an energy gap $\Delta_R$. Hence the dissipation increases faster than $T$ (by a factor of 17 versus 4 between 50 mK and 200 mK). We reason as follows that $\Gamma$ should decrease, consistent with the observed trend in $H_J$. 

At very low $T$, changes caused by increasing $\eta$ compete with purely statistical changes caused by varying $T$. Following Grabert \etal~\cite{Weiss}, the tunneling rate $\Gamma(T,\eta)$ at low $T$ may be expressed as
\be
\Gamma(T,\eta)  = \Gamma_0(\eta) \exp(A(T,\eta)),	\quad\quad \Gamma_0(\eta)\sim \exp(-B(\eta,0)/\hbar), 		
\label{eq:rate}
\ee
where $B(\eta,0)$ is the $\eta$-dependent exponent at $T$ = 0. The $T$-dependent exponent is given by $A(T,\eta) = f(\eta)T^2$ (for Ohmic dissipation).

Although the analytic expression for $B(\eta,0)$ is not known, numerical integration \cite{Chakravarty} shows that, at very large $\eta$, $B(\eta,0)$ is asymptotically linear in $\eta$, but it approaches a constant as $\eta\to 0$. Because they are additive and have opposite signs, $B(\eta,0)$ and $A(T,\eta)$ compete in the exponent of $\Gamma(T,\eta)$. Unfortunately, theory provides only limited guidance to their relative magnitudes. We reason that at low $T$ and small $\eta$, $A(T,\eta)$ is very small compared with $B(\eta,0)$ (in fact Ref. \cite{Weiss} shows that $A(T,\eta)$ vanishes exponentially like $\exp(-\hbar\omega_0/k_BT)$ in the limit $\eta\to 0$, where $\omega_0$ is the characteristic frequency at the well minimum). Hence $B(\eta,0)$ dominates the exponent in $\Gamma(T,\eta)$, i.e. $\Gamma(T,\eta)$ is an overall decreasing function of $\eta$, with negligible dependence on $T$ (except insofar as raising $T$ in our situation causes $\eta$ to increase exponentially). In effect, as we raise $T$ between $T_1$ and $T_2$, $\Gamma$ decreases, consistent with the trend in $H_J(T)$.

Tests of macroscopic quantum tunneling have been carried out in experiments involving Josephson junctions~\cite{Washburn,Devoret,Martinis}, molecular magnets~\cite{Sarachik,Thomas,Barbara99}, and domain wall dynamics~\cite{Barbara93,Wernsdorfer,Rosenbaum}.

A new feature in the present experiment is that both the jump magnitude $\Delta R_{yx}$ and its spatial correlation also provide information on the rapid wall expansion that occurs after tunneling. Because dissipation diverts some of the energy gain, a large dissipation will severely restrict the final bubble size, overdamping the expansion~\cite{Stamp,Chudnovsky}. To us, this provides a persuasive explanation of the behavior of $\Delta R_{yx}(T)$ observed above 120 mK (Fig. \ref{figHJ}C). The magnitude $\Delta R_{yx}$ provides a measure of the size of the spin-up regions after the expansion corresponding to the first jump. In the interval 80 mK $<T<$120 mK, the expansion of the true ground state extends to roughly half the sample. As $T$ rises above 120 mK, dissipation increasingly limits the expansion. Above $T_2$, the final radius (as measured by the jump size) falls below resolution. Even if tunneling persists (as implied by the continued increase in $H_J$ near $T_2$), the final bubbles are far too small to be detected by $R_{yx}$. Raising $T$ in this interval opens a dissipative channel that sharply limits the size of the true ground state bubbles. (Because the activation gap protecting the quantized state is similar in magnitude ($\Delta_R\sim$ 190 mK, Sec. \ref{sec:gap}), we propose that the bulk electrons are the source of dissipation limiting the wall expansion.) 

Above $T_2$, the smooth variations of $R_{yx}$ vs. $H$ (Fig. \ref{figRyxvsT}C) imply a gradual increase of the spin-up domains associated with domain wall motion as in conventional magnets. At all $T$ investigated, this smooth evolution is present as a background process that exists in parallel with the jumps of interest.

\begin{figure}
\includegraphics[width=18cm]{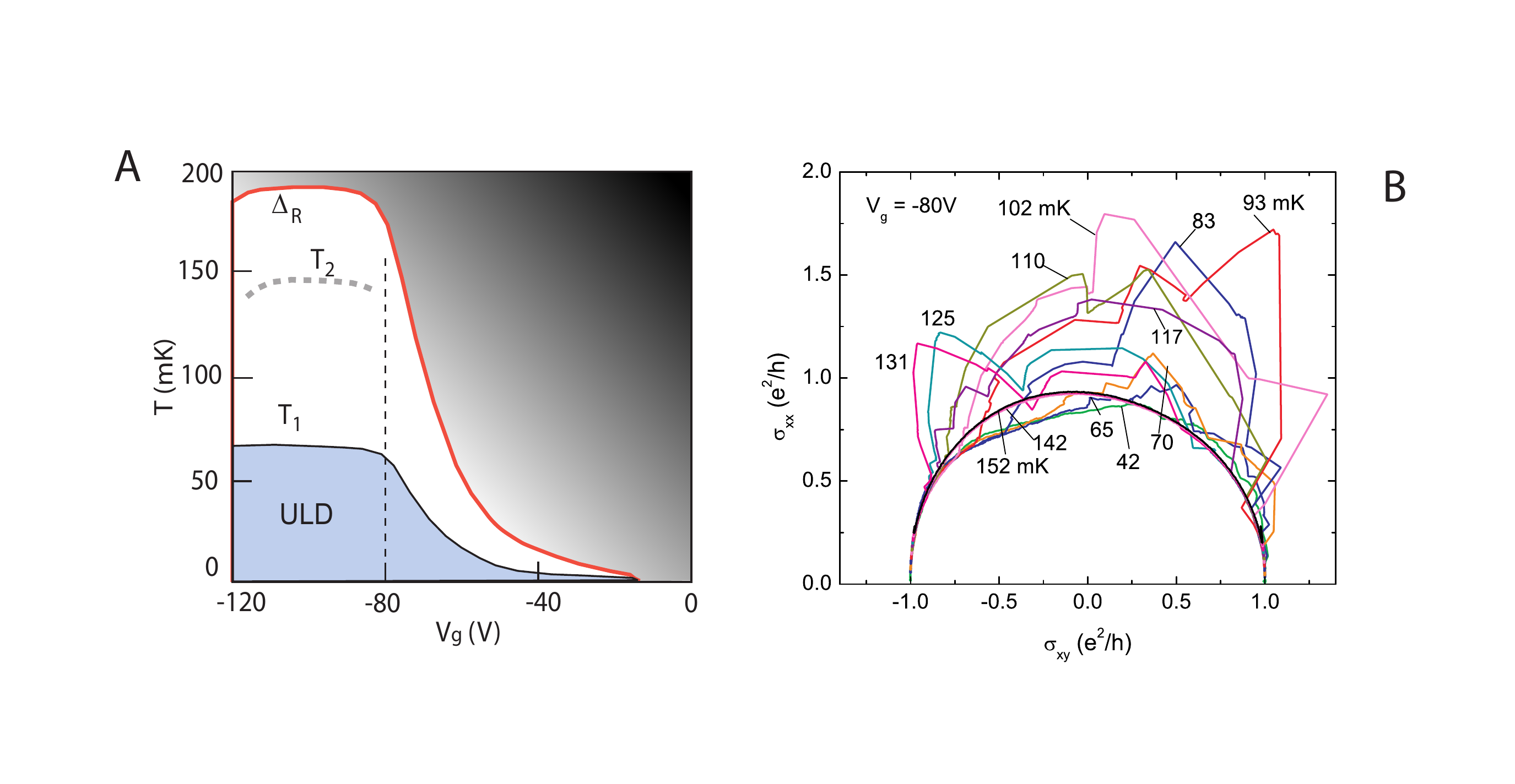}
\caption{\label{figphase}
Crossover boundaries in the $(V_g, T)$ plane (Panel A) and scaling plot of $\sigma_{xx}$ vs. $\sigma_{xy}$ (B). In Panel A, the optimal gating window lies left of the vertical dashed line (-120 $<V_g<$ -80 V). The resistance gap $\Delta_R$, which protects the ultralow dissipation state, equals $\simeq$ 190 mK within this window but falls rapidly outside the window. In the shaded region above $\Delta_R$, the system is dissipative (darker shade indicates larger $R_{xx}$). Jumps in $R_{xx}$ and $R_{yx}$ are observed between the two characteristic temperatures $T_1$ and $T_2$. The curve of $T_1$ decreases steeply outside the window. In the ultralow dissipation region below $T_1$ (ULD), jumps are not observed on our experimental timescales. In Panel B, the scaling plot describes a nearly ideal semicircle above $T_2$ (curves at 142 and 152 mK). Within the interval ($T_1, T_2$), the appearance of jumps kicks the orbits high above the semicircle. This implies violation of the two-parameter scaling behavior~\cite{Pruisken}.
}
\end{figure}

At lower $T$, a major puzzle uncovered is the steady decrease of $\Delta R_{yx}$ to very small steps when $T$ decreases below 70 mK to 10 mK for $V_g$ = -80 V (Figs. \ref{figRyxvsT}B and \ref{figHJ}C). The non-monotonic profile of $\Delta R_{yx}$ vs. $T$ implies that a second process that suppresses the tunneling rate becomes increasingly important at very low $T$. We emphasize that this suppression is confined to the ultra-low dissipation state (the region -120$<V_g<$ -80 V in Fig. \ref{figphase}). Once dissipation is activated by gating, the jumps reappear at 10 mK. As shown in Fig. \ref{figRyxvsT}D, the curve at $V_g$ = -90 V does not show jumps whereas jumps reappear in the curves when $V_g$ is changed to (-40, -10) V (all curves are at 10 mK). 

It is instructive to view this pattern in the $V_g$-$T$ plane (Fig. \ref{figphase}). The reappearance of the jumps for $V_g>$ -40 V implies that the boundary $T_1$ decreases rapidly from 70 mK to $<$10 mK once $V_g$ leaves the optimal gating window. Hence a large gap $\Delta_R$ (which strongly suppresses activation of bulk carriers) favors a large $T_1$ while a small $\Delta_R$ decreases $T_1$ to below 10 mK. The pattern implies that the absence of jumps below $T_1$ is instrinsic to the ultralow dissipation state. 

The observation suggests the following (tentative) picture. Since zero dissipation leads to a steep increase in the correlation length $\xi$ of the spins in the limit $T\to 0$, we reason that the tunneling rate $\Gamma$ will again be suppressed if $N_s$ (the number of spins that tunnel coherently) increases to macroscopic values -- e.g. the size of the whole sample. In the WKB approximation, where $\Gamma \sim\exp(-B/\hbar)$ with $B = 2\int \sqrt{2mU(q)} \;dq$, with $q$ a collective coordinate, a large $N_s$ or large mass $m$ leads to strong suppression of $\Gamma$.

When $\mu$ lies within the minigap, the bulk electrons become negligible in the limit $T\to 0$, so that $\xi$ extends over macroscopic regions and $\Gamma$ becomes very small. Increasing the dissipation by changing $V_g$ to (-40, -10) V serves to reduce the size of the coherent region. Hence jumps can reappear at 10 mK. Although this argument is speculative, the experiment potentially provides a way to measure $\xi$ and its effect on $\Gamma$ in the limit $T\to 0$.

\subsection{Nucleation in Superfluid helium}\label{supercool}
Another known example in which the escape rate increases with decreasing $T$ is the metastable $A$ phase of superfluid $^3$He produced by supercooling below the $AB$ transition. If $^3$He is cooled from the normal state to the superfluid state at a finite pressure $P$ (= 30 bar, say), it first undergoes a second-order transition to the anisotropic $A$ phase at the critical temperature ($T_c \sim$2.3 mK). At a slightly lower $T$ ($\sim$2 mK), a first-order transition to the isotropic $B$ phase occurs. Due to the first-order nature of the $AB$ transition, the $A$ phase is metastable as a supercooled liquid until $\sim 0.7\; T_c\sim$ 1.4 mK when bubbles of the $B$ phase expand to fill the whole sample. 

The Cahn-Hilliard theory (Eq. \ref{eq:Fb}) predicts that, in 3D, $R_c = 2\sigma_{AB}/\Delta F$ and $F^c_b = (2\pi/3) R_c^3\Delta F$, where $\sigma_{AB}$ is the surface tension and $F_B$ the condensation energy of the $B$ phase. Using known quantities, one calculates $F^c_b$ to be extremely large ($10^6\to10^7\; k_BT$ at 0.7$T_c$). The predicted lifetime of the A phase $\tau_A$ ought  to be many times the age of the universe. The observed nucleation temperature posed a deep puzzle until Leggett proposed the theory (dubbed ``baked Alaska'')~\cite{Leggett84} that nucleation is triggered by cosmic rays. The strongly stochastic and often irreproducible nature of the nucleation process made experimental tests difficult. Subsequently, it was shown that $\tau_A$ is reduced by factors of $10-100$ if the superfluid is exposure to gamma rays from a $^{60}$Co source, in strong confirmation of the cosmic-ray theory~\cite{Schiffer2}. Experimentally, $\tau_A(T)$ is observed to decrease steeply with decreasing $T$ with the empirical form $\tau \sim \exp[a(R_c/R_0)^n]$ with $n=3-5$ ($a$ and $R_0$ are fit parameters).

Despite the similarity of the lifetime trend vs. $T$, the 2 experiments are quite different. In $^3$He, the extremely large barrier $F^c_b$ precludes thermal activation as a relevant factor at 1 mK. The $T$ dependence of $\tau_A$ arises because, in the vicinity of the $T_c$ curve, the critical divergence of $\sigma_{AB}(T)$ leads to a divergent increase in $R_c$ (hence of $F^c_b$) as $T$ approaches $T_c$ from below. The primary effect of decreasing $T$ is to strongly decrease the barrier height $F^c_b$. Clearly, this depends on being close to the second-order transition curve $T_c$ (to have a strongly $T$ dependent $R_c$). In addition, we need an intervening first-order ($AB$) transition for supercooling to occur.

In the QAH system, extensive measurements of $R_{ij}$ vs. both $H$ and $T$ (reported here and in Refs. \cite{Xue,Checkelsky,KangWang,Moses,Goldhaber}) have found no evidence for thermodynamic phase transitions (first or second order) below 4 K (the critical temperature for the ferromagnetic transition is in the range 15-30 K). The barrier height in the metastable state (which depends on $H_c$ and the exchange energy) is controlled by $H$, but is essentially $T$-independent below 0.5 K. Because the knob ($H$) that changes the barrier height is independent of the knob that controls the dissipation ($T$), we can infer that the observed trend of $H_J(T)$ reflects the role of dissipation on the escape rate. Both reasons persuade us that the $^3$He supercooling theory is not applicable here.

\subsection{Multiple paths and Network model}\label{network}
The pattern of correlation between Hall signals observed across pairs of Hall contacts separated by 100 $\mu$m (or more) suggests that the jumps strongly perturb the magnetization pattern. As an illustration, we adopt a one-wall model that allows $R_{14,mn}$ to be calculated for assumed positions of the wall using the B\"{u}ttiker formalism (see Supplementary Sec. S4). We calculate that the downstream Hall pair $R_{14,53}$ undergoes a jump, but not the upstream $R_{14,62}$ (Supplementary Fig. S5). $R_{14,23}$ jumps up while $R_{14,65}$ jumps down, in agreement with Figs. 5C and D. Next, we assume that the wall is further displaced to lie between the pair (6,2) and contact 1 at the second jump. The calculated upstream Hall pair undergoes a jump, and the resistance pairs now reverse their jump directions consistent with observation. 
However, the single-wall model is inadequate in many ways. Its biggest difficulty is that the predicted jumps $\Delta R_{yx}$ are always 2 (in units of $h/e^2$) whereas the observed magnitudes are roughly 1 or much smaller. This flaw appears to be intrinsic to the one-wall assumption. 

We generalize to a multidomain picture in which tunneling starts from many seeds distributed across the sample. A natural starting point is the closely related plateau transition in the conventional integer quantum Hall effect (IQHE) which has received considerable attention (see Supplementary Sec. S5)~\cite{Chalker,HoChalker,Ludwig}. As $H$ is varied, the chemical potential $\mu$ crosses the center of a Landau level broadened by the disorder potential $V({\bf x})$. The current is diverted from the dissipationless edge modes to a very small subset of extended but dissipative bulk states at the band center (states away are Anderson localized). In the bulk, the centers of the cyclotron orbits describe chiral trajectories that track the energy contours of $V({\bf x})$. Tunneling occurs between chiral trajectories that get close to each other (this leads to scattering). Hence the transition of $R_{yx}$ from one Hall plateau to the next occurs concurrently with the appearance of dissipation (finite $\sigma_{xx}$). 

The network model studied by Chalker and collaborators (Sec. S4)~\cite{Chalker,HoChalker} provides an elegant way to treat the flow of current through multiply connected paths. The sample is modeled as a regular lattice of corner sharing loops defined by directed links. At the shared corners (the nodes), incoming currents are scattered to one of the two outgoing links with amplitudes parametrized by $\beta$ (Eqs. S10 and S11 in Sec. S4). They find that the localization length $\xi(\beta)$ diverges as $|\beta-\beta_c|^\nu$ with $\nu\sim$2.5 where $\beta = \beta_c$ at maximal dissipation (where $\sigma_{xx}$ peaks).

For the Chern transition in the QAH problem (Eq. S13 in Sec. S4), Wang, Liang and Zhang (WLZ)~\cite{Jing} have established the existence of 3 types of domains with Chern number ${\cal C} = 0, \pm 1$. Identifying the Chern-number domain walls with the energy contours in the IQHE problem, WLZ map the Chern transition to the network model of Chalker \etal, and obtain a divergence in $\xi$ with the same exponent $\nu$. WLZ infer that $\sigma_{ij}$ obeys the two-parameter scaling relationship~\cite{Pruisken}. Evidence for this scaling in the QAH problem has been obtained by plotting $\sigma_{xx}$ vs. $\sigma_{xy}$ in the vicinity of the Chern-number transition~\cite{Checkelsky,KangWang}. Evidence for the $\nu$ = 0 QAH state has also been reported by Yoshimi \etal~\cite{Yoshimi}.

We discuss our results in the context of the foregoing. Just before the first jump in $R_{yx}$ occurs, the sample is predominantly in the ${\cal C} =-1$ state (because $R_{xx}$ is finite and very small, there exist small regions with ${\cal C} = 0$). After the first jump, which involves -- by assumption -- expansion from multiple seeds spread over the sample, the ${\cal C} = 0$ regions define a multiply connected network of conducting paths. The observation that $\Delta R_{yx}$ is close to 1 (but never 2) suggests to us that the converted regions have ${\cal C} = 0$ (this provides support for the 3-domain proposition of WLZ~\cite{Jing}). In this network, connectivity between the chiral modes is mediated by tunneling at the nodes as in the network model (the “microscopic” tunneling between chiral modes occurring at the nodes is to be distinguished from the macroscopic quantum tunneling that triggers the jumps). The bulk conduction is dominated by a few backbone paths (as in percolation networks) that extend across the sample width to give $\Delta R_{yx}\sim 1$. We expect the sudden appearance of a network of regions with ${\cal C} = 0$ to strongly perturb the system away from the two-parameter scaling behavior. This is borne out in an interesting way.

In Fig. \ref{figphase}B, we show plots of $\sigma_{xx}$ vs. $\sigma_{xy}$ at selected $T$ for $V_g$ fixed at -80 V (within the optimal gating window). Above $T_2$ (curves at 142 and 152 mK), the measured $\sigma_{ij}$ trace out a nearly ideal semicircle (the dome). Within the interval $(T_1, T_2)$ the first jump and subsequent ones kick the orbit high above the dome -- the orbit returns only when the transition is nearly complete ($\sigma_{xy}\to e^2/h$). A notable pattern is that the perturbed orbits extend above the dome (the exception is a small sliver inside the dome for pre-jump segments at $T<$ 100 mK). This rich pattern suggests that the two-parameter scaling behavior is valid above $T_2$, but the jumps take the system high above the scaling dome.

Although two-parameter scaling cannot be employed to obtain $\sigma_{xy}$ when jumps occur, it seems possible to calculate numerically $\sigma_{xy}$ and $\sigma_{xx}$ from large networks as carried out for $\xi$ in Refs.~\cite{Chalker,HoChalker}. Calculations clarifying the conditions under which $\Delta R_{yx}$ is smaller than 1 will add valuable insight into what happens at the jumps.

\subsection{Outlook and Conclusions}
The key features uncovered are the large jumps in the Hall resistance of magnitude $\Delta R_{yx}\sim h/e^2$ and in $R_{xx}$ when the QAH system is trapped in the metastable state with ${\cal C} = -1$, within the interval $T_1< T < T_2$. A key finding is that the escape rate is sharply decreased as $T$ increases. From the thermal activation of $R_{xx}$, we have inferred that the increase in dissipation overwhelms any purely thermal effect associated with increasing $T$; the increased dissipation then suppresses the tunneling probability $\Gamma$. Hence the Caldeira-Leggett theory seems to be the most appealing explanation for the jumps. 

Because the transitions involve tunneling of a field $\phi({\bf x},t)$ with many degrees of freedom (as opposed to one collective degree of freedom), the present experiment uncovers additional features that invite further investigation. The most interesting seems to be the existence of a temperature scale $T_1$ below which jumps are not observed within the experimental timescale, which we tentatively identify with another effect of dissipation. As shown in Fig. \ref{figphase} 
the ultralow dissipation state is protected by a small energy gap $\Delta_R$ from thermal excitation of bulk electrons. The latter opens up a dissipative channel that strongly affects the tunneling process. Above $T_2$, the dissipation strongly limits the expansion of the bubble after tunneling occurs.

In a recent scanning nano-SQUID investigation of the magnetization reversal in a sample closely similar to ours, the existence of very small (several 10 nm) spin-up regions was directly observed at $T$ down to 300 mK~\cite{Zeldov}. Strikingly, these small regions are static and hence incompatible with conventional domain wall motion. We suggest that the static spin-up regions are bubbles whose post-tunneling expansion has been halted by overdamping (at 300 mK, they are too small to be resolved in $R_{yx}$). Extending the SQUID experiments to below 100 mK (where $\Delta R_{yx}\sim 1$) will allow a direct comparison with the transport results. We expect that the static size of the spin-up regions will increase rapidly to macroscopic lengths (100 $\mu$m) as $T$ is lowered from $T_2$ to 100 mK. A related experiment revealing the coincidence of superparamagnetism and perfect quantization in the QAH problem was recently reported~\cite{Grauer}.

Investigation of the correlations between jumps can be feasibly extended to a large number ($\sim$20) of contacts to provide a more detailed map of the spatial displacement of the wall. 

The role of dissipation in macroscopic quantum tunneling (for e.g. in determining $N_s$) remains a challenge both for theory and experiment. The quantum anomalous Hall system, in which the chemical potential, dissipation, field strength and temperature can be varied independently, provides a powerful system to explore these issues quantitatively.

\section{MATERIALS AND METHODS}
Samples were grown by molecular beam epitaxy (MBE) using thermally heated single elemental (at least 5N purity) sources. The SrTiO$_3$ substrates were annealed in an oxygen environment between 925 C to 965 C for about 2.5 hours and were checked after annealing with atomic force microscopy for an atomically ordered surface with an RMS roughness of $<$0.3 nm before growth. Substrates were indium mounted on the MBE sample holder and outgassed at ~500 C for an hour in the MBE chamber before growth. Cr doped (Bi,Sb)$_2$Te$_3$ was grown at ~ 240 C as measured by a pryometer with a growth rate of about 0.5 QL/min. After growth, the sample was allowed to cool to room temperature and was capped with about 3 nm of Al. The sample was then exposed to atmosphere to allow the Al capping layer to oxidize, forming Al$_2$O$_3$. Finally, the sample was heated in an argon environment to melt the indium in order to free the sample from its substrate. The remaining indium on the back of the substrate is then used for the back gate contact. For further details, see Kandala \etal~\cite{Nitin}

In the experiment, the Hall bar (width 0.5 mm and length 1 mm) was made by scratching the thin film with a sharp needle controlled by programmable stepper motors. The Ohmic contact is made by firmly pressing tiny pieces of Indium onto the contact pads. The pressing action is believed to help the Indium penetrate through the Al$_2$O$_3$ capping layer and form a good contact with the underlying Cr doped (Bi,Sb)$_2$Te$_3$ layer. The measurements were performed with the sample immersed in the $^3$He-$^4$He mixture of a top-loading wet dilution refrigerator (Oxford Instruments TLM400). To filter out room-temperature radiation, three filter stages were added to the circuit. At the room-temperature end, an LC $\pi$ filter provided over 80 dB attenuation above 100 MHz. Two cryogenic filters were installed in the mixing chamber. One was a powder filter that filters out radiation above 1 GHz while the second was an RC filter with a cutoff frequency of ~300 KHz. The overall time constant of the whole circuit was $\sim$10 ms. Nearly all the results reported were taken using a Keithley 6221 precision current source and a Keithley 2182 nanovoltmeter operating in delta mode (with overall time constant of $\sim$1 s). The relative accuracy of the resistance measurements was confirmed to be better than 0.1$\%$ using a 100 ppm precision resistor. The exceptions (Figs.  2D, 4, 5 and 6 of the main text) were performed using a Stanford Research SR830 lock-in amplifier at frequency ~2 Hz, with the signal pre-amplified by a Stanford Research SR560 and/or PAR113 preamplifiers (time constant set at 2 s in Figs. 4, 5 and 6, and 5 s in Fig. 2D). The excitation current was 1 nA in both methods. 

For the high-speed measurements of the jump duration (Fig. \ref{figTime}), we applied a DC current of 2.5 nA. The room-temperature filter was removed. The voltage signal was amplified by the SR560 preamplifier and then digitized by a Zurich Instrument HF2 lock-in amplifier at a sampling rate of 6.4 KS/s (S= sampling). For these measurements the field-sweep rate was reduced to 0.1 mT/min.

\section{REFERENCES AND NOTES}

\vspace{5mm}\noindent
{\bf Supplementary Information} is available in the online version of the paper.

\vspace{5mm}\noindent
{\bf Acknowledgements:} {\bf Funding} We acknowledge support from DARPA SPAWAR (N66001-11-1-4110) and a MURI award for topological insulators (ARO W911NF-12-1-0461). N.P.O. acknowledges support by the Gordon and Betty Moore Foundation’s EPiQS Initiative through Grant GBMF4539. We thank David Goldhaber-Gordon, Jing Wang, Yayu Wang and Shou-Cheng Zhang for valuable discussions.\\

\vspace{5mm}\noindent
{\bf Author Contributions:} M.H.L., N.S., A.Y. and N.P.O. designed the experiment. A.R.R. and N.S. grew the samples. M.H.L., W.W. and A.K. performed the measurements. M.H.L., W.W., J.L., A.Y. and N.P.O. carried out the analysis and modelling of the results. M.H.L. and N.P.O. wrote the manuscript with inputs from all authors. 

\vspace{5mm}\noindent
{\bf Competing interests:} The authors declare no competing financial interests. Correspondence and requests for materials should be addressed to M.H.L. (minhaol@princeton.edu) and N.P.O. (npo@princeton.edu).

%%%%%%%%%%%%%%%%%%%%%%%%

\newpage

%%%%%%%%%%%%%%%%%%%%%%%%%%%%%%%%%%
%%%%%%%%%%%%%%%%%%%%%%%%%%%%%%%%%%
%%%%%%%%%%%%%%%%%%%%%%%%%%%%%%%%%% FIGURE 1

%%%%%%%%%%%%%%%%%%%%%%%%%%%%%%%%%%
%%%%%%%%%%%%%%%%%%%%%%%%%%%%%%%%%%
%%%%%%%%%%%%%%%%%%%%%%%%%%%%%%%%%% FIGURE 2

%%%%%%%%%%%%%%%%%%%%%%%%%%%%%%%%%%
%%%%%%%%%%%%%%%%%%%%%%%%%%%%%%%%%%
%%%%%%%%%%%%%%%%%%%%%%%%%%%%%%%%%% FIGURE 3

%%%%%%%%%%%%%%%%%%%%%%%%%%%%%%%%%%
%%%%%%%%%%%%%%%%%%%%%%%%%%%%%%%%%%
%%%%%%%%%%%%%%%%%%%%%%%%%%%%%%%%%% FIGURE 4

%%%%%%%%%%%%%%%%%%%%%%%%%%%%%%%%%%
%%%%%%%%%%%%%%%%%%%%%%%%%%%%%%%%%%
%%%%%%%%%%%%%%%%%%%%%%%%%%%%%%%%%% FIGURE 5

%%%%%%%%%%%%%%%%%%%%%%%%%%%%%%%%%%
%%%%%%%%%%%%%%%%%%%%%%%%%%%%%%%%%%
%%%%%%%%%%%%%%%%%%%%%%%%%%%%%%%%%% FIGURE 6

%%%%%%%%%%%%%%%%%%%%%%%%%%%%%%%%%%
%%%%%%%%%%%%%%%%%%%%%%%%%%%%%%%%%%
%%%%%%%%%%%%%%%%%%%%%%%%%%%%%%%%%% FIGURE 7

%%%%%%%%%%%%%%%%%%%%%%%%%%%%%%%%%%
%%%%%%%%%%%%%%%%%%%%%%%%%%%%%%%%%%
%%%%%%%%%%%%%%%%%%%%%%%%%%%%%%%%%% FIGURE 8

%%%%%%%%%%%%%%%%%%%%%%%%%%%%%%%%%%
%%%%%%%%%%%%%%%%%%%%%%%%%%%%%%%%%%
%%%%%%%%%%%%%%%%%%%%%%%%%%%%%%%%%% FIGURE 9

\end{document}